\documentclass[prl,twocolumn,showpacs,preprintnumbers]{revtex4}
\usepackage{graphicx}
\usepackage{dcolumn}

\begin{document}

\title{Deformation Electron-Phonon Coupling \\
in Disordered Semiconductors and Nanostructures}

\author{ A. Sergeev}
\email{asergeev@eng.buffalo.edu} \affiliation{ Research
Foundation, University at Buffalo, Buffalo, New York 14260}
\author{M. Yu. Reizer} \affiliation{5614 Naiche Rd.
Columbus, Ohio 43213}
\author{V. Mitin} \affiliation{Electrical Engineering
Department, University at Buffalo, Buffalo, New York 14260}


\begin{abstract}
We study the electron-phonon relaxation (dephasing) rate in
disordered semiconductors and low-dimensional structures. The
relaxation is determined by the interference of electron
scattering via the deformation potential and elastic electron
scattering from impurities and defects. We have found that in
contrast to the destructive interference in metals, which results
in the Pippard ineffectiveness condition for the electron-phonon
interaction, the interference in semiconducting structures
substantially enhances the effective electron-phonon coupling. The
obtained results provide an explanation to energy relaxation in
silicon structures.

\end{abstract}
\pacs{PACS   numbers: 72.10.D}

\maketitle

Elastic electron scattering from impurities and defects
drastically changes the electron-electron and electron-phonon
(e-ph) interaction and modifies temperature dependencies of the
relaxation/dephasing rate. As a result of diffusive motion of
electrons, the electron-electron interaction is significantly
enhanced in bulk and low-dimensional conductors \cite{AA}. Recent
theoretical \cite{Chow,Reizer} and experimental \cite{Chow,
Fletcher} studies have shown that the piezoelectric e-ph coupling
is also enhanced in semiconductors with short electron mean free
path. Effects of disorder on the deformation e-ph coupling are
more complicated. In impure metals, the deformation coupling
originates from "pure" electron-phonon scattering, electron
scattering from vibrating impurities, and various interference
processes. If electron scatterers vibrate in the same way as host
atoms, the destructive interference of scattering mechanisms
\cite{Sch, RS} results in the Pippard ineffectiveness condition
\cite{P}, which means suppression of the e-ph relaxation. In this
case, at low temperatures the relaxation rate modifies from
$T^3$-dependence in the pure materials to $T^4$-dependence in the
impure metals. However, even small amount of static scatterers
(e.g. tough boundaries) or incomplete drag of impurities and
defects increases the e-ph relaxation \cite{SM}.
Disorder-suppressed relaxation is observed in disordered metallic
films \cite{Ger, Maas}, while alloys commonly demonstrate the
disorder-enhanced relaxation with $T^2$-dependence of the
relaxation rate \cite{Lin}.

Recently, there has been significant interest to the electron
relaxation in disordered semiconductors and structures,  where the
electron relaxation is determined by electron-phonon scattering
via the deformation potential (DP). The relaxation rate has been
measured in Si crystals containing $\delta$(Sb)-layer
\cite{Kashirin,Komnik} and in Si films \cite{Kivinen}.
Experimental results, including $T^4$-dependence of the electron
relaxation rate, were associated with the Pippard ineffectiveness
condition, obtained for metals. However, in the temperature ranges
investigated in Refs. \cite{Kashirin,Komnik} and \cite{Kivinen},
DP is strongly screened and the relaxation rate in pure 2D and
quasi-2D-structures follows to $T^5$-dependence \cite{CCC,AAA}.
Therefore, the ineffectiveness would result in the $T^6$, rather
than the $T^4$-dependence.

It is not surprising that the theory developed for metals
\cite{Sch,RS,P} fails to describe semiconductors. Indeed, DP in
metals and semiconductors has different nature \cite{Gant}. In
metals, DP is associated with electron gas compressibility, while
in semiconductors this contribution is negligible due to small
carrier concentrations. DP in semiconductors results mainly from a
shift of the conduction-band edge under the deformation, while in
metals such contribution is small because of strong screening. It
is important that DP has different tensor structures in metals and
semiconductors \cite{Met,Gant}, and this difference clearly
manifest itself even in weakly disordered conductors \cite{SRM}.
Here we show that the tensor structure of DP plays a crucial role
in kinetics  of strongly disordered conductors: in contrast to the
destructive interference in metals, the electron-phonon-impurity
interference in semiconducting structures substantially enhances
e-ph coupling.

Here we report results on the e-ph relaxation in disordered bulk
semiconductors, two-dimensional electron structures, and
multi-channel one-dimensional conductors interacting with 3D
phonons. Effects of disorder are described by the dimensionless
parameter $q l$, where $q$ is the characteristic momentum
transferred to the electron due to e-ph scattering, $l=v_F \tau$
is the electron mean free path due to scattering from impurities,
$v_F$ is the Fermi velocity. In the impure limit, $q l \ll 1$, a
phonon interacts with an electron that diffuses in the interaction
region, $L \sim 1/q \gg l$. In bulk conductors, $q$ is the
wavevector of a thermal phonon, $q_T=T/u$ ($u$ is the sound
velocity), and the crossover to the impure limit occurs at $T \sim
u/l $.

In low-dimensional conductors, the characteristic momentum $q$ is
determined by the phonon wavevector component $q_\parallel$, which
is parallel to the conductor. In two-dimensional systems,
$q_\parallel$ is of the order of $q_T$ and, as well as in bulk
semiconductors, the crossover is described by the parameter $q_T
l=Tl/u$. In 1D channels the transferred momentum $q_\parallel$ is
$\sim(u/v_F)q_T$ and effects of disorder become important at
significantly higher temperatures, $ T \tau \sim q_\parallel l <
1$. The same parameter $T\tau$ describes modification of the
electron-electron interaction \cite{AA}.

Investigating the electron energy relaxation, we focus our
attention on the time scale much longer than the electron momentum
relaxation time. In this time domain, electron-phonon kinetics is
described by the angle-averaged electron and phonon distribution
functions, $n_\epsilon $ and $N_\omega $. We consider interference
processes, which are characterized by the momentum transfer much
smaller than the Fermi momentum. In this case, the interference of
electron-phonon and electron-impurity scattering is taken into
account by the electron self-energy diagram shown in Fig. 1. The
corresponding electron-phonon collision integral is \cite{RS,SM}
\begin{eqnarray}\label{I}
I_\epsilon &=& -4 \tau \int {d{\bf q} d\omega \over (2\pi)^4} \
{\gamma^2 \over |\epsilon^R_n|^2}  \Im D^R(q, \omega) \ \Re
{\zeta_n \over 1-\zeta_n} \nonumber \\ &\times& N_\omega
n_\epsilon (1-n_{\epsilon+\omega})-(1+N_\omega) (1-n_\epsilon)
n_{\epsilon+\omega}, \ \
\end{eqnarray}
where $D^R(q,\omega)$ is the phonon Green function
\begin{equation}\label{D}
 D^R({\bf q},\omega) =
(\omega-\omega_{\bf q} + i0)^{-1}+ (\omega+\omega_{\bf q} +
i0)^{-1},
\end{equation}
$\gamma = D_0 ({\bf q}\cdot{\bf e})/(2 \rho
 \omega_q)^{1/2}$ is the vertex of the electron-phonon interaction,
$D_0$ is the constant of DP, and ${\bf e}$ is the phonon
polarization vector.

In the collision integral $I_\epsilon$, $\zeta_n$ is an integral
over the impurity-averaged electron Green functions \cite{RS},
\begin{eqnarray}\label{zeta}
\zeta_n = {1 \over \pi \nu_n \tau} \int {d{\bf p}\over (2\pi)^n}
G^A({\bf p}, \epsilon) G^R({\bf p}+{\bf q}, \epsilon+ \omega)
\nonumber \\ = \cases{{\displaystyle \ \ \ {\arctan(ql)\over
(ql)}} &  $n=3$, \cr {\displaystyle {1\over \sqrt{1+(q_\|l)^2}}} &
$ n=2$, \cr {\displaystyle {1-i\omega \tau \over (1- i\omega
\tau)^2 + (q_\parallel l)^2}} & $n=1$, }
\end{eqnarray}
where $\nu_n$ is the two-spin electron density of states in
$n$-dimensional electron system. We limited our consideration by
the condition $ql
> u/v_F \sim 10^{-2}$,
which allows us to put $\omega=0$ in $\zeta_3$ and $\zeta_2$.

The screening of DP is described by the dielectric function
$\epsilon_n^{R}(q, \omega)$. Further calculations show that in the
3D and 2D electron systems the characteristic frequencies $ \omega
\sim T $ are small compared with $D q^2$ ($D$ is the diffusion
coefficient). In this limit the dielectric function is
\begin{eqnarray}\label{eps}
\epsilon_n^{R}(q) = \cases{ {\displaystyle {1 + (\kappa_3/q)^2, \
\ \kappa_3^2 = 4\pi e^2 \nu_3}} & \ 3D; \cr {\displaystyle { 1 +
(\kappa_2/q_\|), \ \ \ \kappa_2 = 2\pi e^2 \nu_2}} & \ 2D.}
\end{eqnarray}
For 1D conductors we should take into account the dynamical
character of electron screening. If $q_\|r \gg 1$ ($r$ is the
conductor radius), the dielectric function is
\begin{eqnarray}\label{eps1}
\epsilon^R= 1+ e^2\nu_1 \ \ln {1 \over (q_\|r)^2} \cdot {(q_\|l)^2
\over  (q_\|l)^2 -(\omega\tau)^2-i \omega \tau}.
\end{eqnarray}

The e-ph relaxation rate is calculated as a variation of the
collision integral $\tau_{e-ph}^{-1} = -\delta I_{e-ph}/ \delta
n_e $. In equilibrium, $N_\omega=N^{eq}_\omega(T)$ and
$n_\epsilon=n^{eq}_\epsilon(T)$, and the relaxation rate of
electrons at the Fermi surface ($\epsilon = 0$) is
\begin{equation}\label{tau}
{1\over \tau_{e-ph}} =  4\tau \int {d{\bf q} \over  (2 \pi)^3 }
{\gamma^2 \over |\epsilon^R|^2} \ (N^{eq}_{\omega_q} +
n^{eq}_{\omega_q}) \ \Re {\zeta_n(\omega_q) \over
1-\zeta_n(\omega_q)}.
\end{equation}
We also calculate the heat flux from hot electrons with the
temperature $\theta$ to phonons with the temperature $T$. The heat
flux may be presented through the energy control function $F(T)$
as
\begin{eqnarray}
 P(\theta, T) = \nu_n \int d\epsilon \ \epsilon I_\epsilon
(\theta, T) \ = F(\theta) - F(T), \\ \label{F}  F(T) =  4\tau
\nu_n \int {d {\bf q} \over (2\pi)^3 } \ {\gamma^2 \over
|\epsilon^R|^2} \ \omega_q^2 N_\omega^{eq} \ \Re
{\zeta_n(\omega_q) \over 1-\zeta_n(\omega_q)}.
\end{eqnarray}

First we calculate the relaxation rate in a bulk semiconductor.
Substituting $\zeta_3$ (Eq. \ref{zeta}) and $\epsilon^R_3(q)$ (Eq.
\ref{eps}) into Eq. \ref{tau} we find
\begin{eqnarray}\label{bulk}
{1\over \tau_{e-ph}}&=&  {D_0^2 \nu_3 \over \rho u^2} {T^3 \over
(p_F u)^2} \ F(q_Tl, q_T/\kappa_3), \ \
\
\\ F(y,z)&=&  \int_0^\infty dx \ x^2 {xy
\arctan(xy) \over xy-\arctan(xy)} \nonumber \\ &\times& \biggl(
{(xz)^2 \over (xz)^2 +1 } \biggr)^2 (N_x^{eq}+n_x^{eq}).
\end{eqnarray}
These formulas in limiting cases are summarized in Tab. I. In the
pure limit, $Tl/u \gg 1$, we reproduce well-known results
\cite{Gant}: in the case of weak screening, $T>u \kappa_3$, the
relaxation rate is proportional to $T^3$; for screened DP, $T < u
\kappa_3$, the relaxation rate changes as $T^7$. In the impure
limit the relaxation rate is proportional to $T^2/l$ for
unscreened DP and to $T^6/l$ for the screened DP. Thus, contrary
to the Pippard ineffectiveness condition in metals \cite{P, Sch,
RS}, the relaxation rate in semiconductors is enhanced by a factor
of $u/(Tl)$ due to elastic electron scattering.  The energy
control function may be estimated as $F(T) \simeq C_e T
/\tau_{e-ph}$, where $C_e$ is the electron heat capacity. In Tab.
I we present F(T) with exact coefficients, because measurements of
F(T) are widely used to obtained $D_0$.

Now we consider the e-ph relaxation in two-dimensional electron
gas. Using Eqs. \ref{zeta} and \ref{eps}, we find that the
relaxation rate (Eq. \ref{tau}) may be presented as
\begin{eqnarray}\label{2D}
{1\over \tau_{e-ph}}&=&  {D_0^2 T^3  \over 2 \pi^2 \rho v_F u^4} \
\Phi(q_Tl, q_T/\kappa_2), \ \ \ \\  \Phi(y,z)&=& \int_0^\infty dx
\ x^2 \int_0^{\pi/2} d\theta {xy  \over \sqrt{1+(xy\sin
\theta)^2}-1} \nonumber
\\ &\times& {(xz)^2 \sin^3 \theta \over (xz \sin \theta +1)^2 }
(N_x^{eq}+n_x^{eq}).
\end{eqnarray}
These formulas in limiting cases are summarized in Tab. II. In the
pure limit, we reproduce well-known results \cite{CCC}. In the
temperature range $T \gg \kappa_2 u$, where DP is weakly screened,
the relaxation rate is proportional to $T^3$; for strongly
screened DP the relaxation rate changes as $T^5$. In the impure
limit, in the case of weak screening, the relaxation rate is
proportional to $T^2 \ln T$ and inversely proportional to $l$. At
low temperatures, where DP is strongly screened, the relaxation
rate is proportional to $T^4/l$. Thus, in heterostructures elastic
electron scattering significantly enhances the e-ph interaction.

Finally, we consider the e-ph interaction in the multichannel 1D
system. Channels may be associated with wires, shells, and
electron subbands. Variations of the multichannel model are
applied to one-dimensional organic conductors, CuO-chains in
high-T$_c$ superconductors, and multi-wall carbon nanotubes
\cite{Mishch}. For simplicity we consider identical channels and
neglect the Coulomb interaction between channels. We suggest that
electrons are scattered between channels and interchannel
scattering prevails over backscattering in the same channel, so
the system is in the conducting state. Electron-phonon scattering
keeps an electron in the same channel and, therefore, it is
screened by electrons in this channel. Using Eqs. \ref{zeta} and
\ref{tau}, we find that without screening the relaxation rate in
the pure conductor is given by
\begin{eqnarray}\label{1Dpure}
{1\over \tau_{e-ph}}= {7\zeta(3)\over 8 \pi} {D_0^2 T^3 \over \rho
v_F u^4}.
\end{eqnarray}
Calculating the integrant in Eq. \ref{tau} in the general case,
note that for a 1D conductor $q_\parallel= q\cos \phi= qx$ ($\phi$
is the angle between ${\bf q}$ and a wire ) and within the
logarithmic accuracy the integral over the direction of ${\bf q}$
is given by
\begin{eqnarray}\label{int2}
\int_0^1 {dx\over 2}  {(qlx)^2 \over [(1-2 e^2 \nu_1 \ln q r x)(q
l x)^2-(\omega \tau)^2]^2+(\omega \tau)^2} \nonumber
\\= {1\over 2ql (1-2 e^2 \nu_1 \ln
q_c r)^{3/2}}
 \cases{ {\displaystyle {\pi / 2}} & $
\omega \tau \gg 1$; \cr {\displaystyle {\pi / \sqrt{8 \omega \tau}
}} & $ \  \omega \tau \ll 1 $,}
\end{eqnarray}
Eq. \ref{int2} shows that the crossover to the impure limit is
described by the parameter $\omega \tau$, which is of the order of
$T\tau$. In the impure limit, $T \tau \ll 1$, the characteristic
value of the transferred momentum $q_c$ is $l^{-1}
\sqrt{\omega\tau} /(l\sqrt{1+e^2 \nu_1}) \simeq l^{-1} \sqrt{T
\tau} /(l\sqrt{1+e^2 \nu_1})$. In this case, the relaxation rate
and energy control function for the one-dimensional multichannel
conductor are
\begin{eqnarray}\label{1Dimpure}
{1\over \tau_{e-ph}}={ 3 \bigl(8-\sqrt{2}\bigr) \zeta(5/2) \over
64 \sqrt{2\pi}\biggl(1- e^2 \nu_1 {\displaystyle \ln (q_c r)^2
\biggr)^{3/2}} }  {D_0^2 T^{5/2} \over \sqrt{\tau} \rho v_F u^4}
\nonumber \\ F(T)={ 105 \zeta(9/2)\over 128 \sqrt{\pi} \biggl(1-
e^2 \nu_1 {\displaystyle \ln (q_c r)^2 \biggr)^{3/2} }} {D_0^2
\nu_1 T^{9/2} \over \sqrt{\tau} \rho v_F u^4}.
\end{eqnarray}
As seen from Eq. \ref{1Dimpure}, screening substantially changes
values of $\tau_{e-ph}$ and $F(T)$, but just weakly affects the
temperature dependencies. Compare Eqs. \ref{1Dpure} and
\ref{1Dimpure}, we find that in the impure limit the electron
phonon interaction is enhanced by the factor of $1/ \sqrt
{T\tau}$.

The electron-phonon-impurity interference in metals and
semiconductors may be qualitatively understood in the following
way. First, elastic electron scattering effectively averages DP
over the Fermi surface. Second, the diffusive motion holds an
electron in the interaction region and increases the interaction
time. In metals the Fermi surface average of the deformation
potential equals to zero \cite{Met,Gant}.  As a result of this
averaging the effective e-ph vertex is substantially decreased
(see Ref. \cite{RS}). In metals this effect prevails over the
modification of the interaction time and strongly suppresses the
e-ph relaxation. In semiconductors, DP weakly depends on the
electron momentum and the DP tensor is usually approximated by a
constant. Therefore, elastic scattering in semiconductors enlarges
the interaction time, which in turn enhances the e-ph relaxation.

Recently the e-ph relaxation rate has been directly measured in 2D
electron gas in Si with MBE-grown Sb $\delta$-layer
\cite{Kashirin,Komnik}. Because of lack of the theory for
semiconducting materials and structures, the observed
$T^4$-dependence was associated with the Pippard concept of the
ineffectiveness of the e-ph interaction. According to our results,
the $T^4$-dependence in 2D structures originates from {\it
disorder-enhanced} screened DP coupling (see Tab. II). Analogous
data with $T^4$-dependence have been obtained in heavily doped
quasi-two-dimensional Si films at subKelvin temperatures
\cite{Kivinen}. Note, that e-ph relaxation rate is often evaluated
from the electron dephasing rate. Such data also give evidence in
favor of significant enhancement of e-ph coupling in disordered
semiconductors. For example, in 3D  Si:P layers with $l \sim 5 $
nm the relaxation time at 4.2 K was found to be ~ 10 ps
\cite{Klap} which is significantly shorter than that in pure
materials.

To conclude, we calculate the e-ph relaxation rate in disordered
semiconductors (Eq. \ref{bulk}, Tab. 1), two-dimensional (Eq.
\ref{2D}, Tab. 2) and one-dimensional (Eq. \ref{1Dimpure})
semiconducting structures. Our results show that the e-ph
relaxation is strongly enhanced due to disorder. The research was
supported by the ONR and MRCAF grants.

\pagebreak

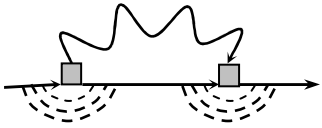
\begin{figure}
\caption{Electron self-energy diagram. Wavy line stands for e-ph
scattering, a dotted line stands for to elastic electron
scattering from random potential, and a straight line stands for
the electron Green function.}
\end{figure}

\begin{table*}
\caption{Electron-phonon energy relaxation time and energy control
function in a bulk semiconductor.}
\begin{ruledtabular}
\begin{tabular}{ccccc}
 &\multicolumn{2}{c}{$T>u \kappa_3$ (weak screening)} &\multicolumn{2}{c}
 { $ T<u\kappa_3$ (strong screening)} \\
 \empty & $T>u/l$ & $T<u/l$ & $T>u/l$ & $T<u/l$ \\[6pt] \hline \\[3pt]
  ${\displaystyle  \tau_{e-ph}^{-1} }$ &
$ {\displaystyle {7 \pi \zeta(3) \over 4}{D_0^2 \nu_3\over \rho
u^2} {T^3\over (p_Fu)^2} }$ & ${\displaystyle {3 \pi^2 \over
4}{D_0^2 \nu_3\over \rho u^2} {T^2 \over p_F^2 lu}}$
&${\displaystyle {5715 \pi \zeta(7) \over 8}{D_0^2 \nu_3\over \rho
u^2} { T^7 \over p_F^2 \kappa_3^4 u^6}}$ & ${\displaystyle {3
\pi^6 \over 4}
 {D_0^2 \nu_3\over \rho u^2} {  T^6 \over p_F^2 \kappa_3^4 l u^5 }} $
\\[12pt]
 $F(T)$ &
  ${\displaystyle 6 \pi \zeta(5) {D_0^2 \nu_3^2\over \rho u^2}
{T^5\over (p_F u)^2} }$ \ & ${\displaystyle { \pi^4 \over
10}{D_0^2 \nu_3^2\over \rho u^2} {T^4 \over p_F^2 l u }}$ &
${\displaystyle 10080 \pi \zeta(9) {D_0^2 \nu_3^2 \over \rho u^2}
{  T^9 \over p_F^2 \kappa^4 u^6}}$ & ${\displaystyle {4 \pi^8
\over 5}{D_0^2 \nu_3^2 \over \rho u^2} {  T^8 \over p_F^2
\kappa_3^4 l u^5}} $  \\
\end{tabular}
\end{ruledtabular}
\end{table*}

\begin{table*}
\caption{Electron-phonon energy relaxation time and energy control
function in two-dimensional electron structures.}
\begin{ruledtabular}
\begin{tabular}{ccccc}
 &\multicolumn{2}{c}{$T>u \kappa_2$ (weak screening)} &\multicolumn{2}{c}
 { $ T<u\kappa_2$ (strong screening)} \\
 \empty & $T > u/l$ & $T<u/l$ & $T>u/l$ & $T<u/l$ \\[6pt] \hline \\[3pt]
  ${\displaystyle  \tau_{e-ph}^{-1} }$ &
${\displaystyle {7\zeta(3)\over 4 \pi}{D_0^2 T^3\over \rho v_F
u^4} }$ & ${\displaystyle {D_0^2  T^2 \over \rho \ l v_F u^3 } \
\ln{T \over \kappa_2 u}}$ &${\displaystyle {93\zeta(5)\over
8\pi}{D_0^2 T^5\over \rho \kappa_2^2 v_F u^6}} $ & ${\displaystyle
{\pi^2\over 4}{D_0^2 T^4\over \rho \kappa_2^2 \ l v_F u^5} }$
\\[12pt]
 $F(T)$ & ${\displaystyle {6 \zeta(5) \over \pi} {D_0^2 \nu_2 T^5 \over \rho v_F u^4}}$ \ & $ {\displaystyle {2 \pi^2\over 15} {D_0^2 \nu_2 T^4\over
\rho \ l v_F u^3 } \  \ln{T \over \kappa_2 u}} $ & ${\displaystyle
{90 \zeta(7)\over \pi} {D_0^2 \nu_2 T^7 \over \kappa_2^2 v_F
u^6}}$ & ${\displaystyle {8 \pi^4\over 63}{D_0^2 \nu_2 T^6 \over
\rho \kappa_2^2 \ l v_F u^5}}$  \\
\end{tabular}
\end{ruledtabular}
\end{table*}

\end{document}